\documentclass{mn2e}

\usepackage{times}
\usepackage{graphicx}

\newif\ifAMStwofonts
\def\til{$\sim$}
\def\msun{${\rm M}_{\odot}$}

\def\deg{$^{\circ}$}

\def\th{\thinspace}

\def\hb{H$\beta$\space}
\def\hg{H$\gamma$\space}

\def\he{He\thinspace{\sc i}\thinspace{$\lambda$}4471\space}
\def\heg{He\thinspace{\sc i}\thinspace{$\lambda$}4388\space}
\def\heh{He\thinspace{\sc i}\thinspace{$\lambda$}4713\space}

\def\heii{He\thinspace{\sc ii}\thinspace{$\lambda$}4686\space}

\def\helium{He\thinspace{\sc i}\space}
\def\hee{He\thinspace{\sc ii}\space}
\def\heu{He\thinspace{\sc ii}\thinspace{$\lambda$}1640\space}
\def\hi{H\thinspace{\sc i}\space}

\def\hhee{He\thinspace{\sc ii}}
\def\hhe{He\thinspace{\sc i}\thinspace{$\lambda$}4471}
\def\hheg{He\thinspace{\sc i}\thinspace{$\lambda$}4388}

\title[X2127+119 in M15: Evidence for a Stripped-Giant Companion]
      {The X-Ray Binary X2127+119 in M15: Evidence for a Very-Low-Mass, Stripped-Giant Companion}
\author[L. van Zyl et al.]
       {L. van Zyl$^{1,2,}$\thanks{E-mail: lvz@astro.keele.ac.uk}, P.A. Charles$^{2,3}$, S. Arribas$^{4}$, T. Naylor$^{1,5}$, E. Mediavilla$^{6}$, C. Hellier$^{1}$ \\
       $^1$Astrophysics Group, School of Chemistry and Physics,
           Keele University, Staffordshire, ST5 5BG\\
       $^2$Department of Astrophysics, Oxford University, Keble Road, 
           Oxford OX1 3RH\\
       $^3$Department of Physics \& Astronomy, University of
           Southampton, Southampton SO17 1BJ\\
       $^4$Space Telescope Science Institute. 3700 San Martin Drive, Baltimore,
           MD21218, USA. On assignment from the Space Telescope\\
           $\;\;$Operations
           Division of the European Space Agency (ESA). On leave from the
           Instituto de Astrofisica de Canarias (IAC) and from\\
           $\;\;$the Consejo
           Superior de Investigaciones Cientificas (CSIC).\\
       $^5$School of Physics, University of Exeter, Stocker Road, Exeter EX4 4QL\\
       $^6$Instituto de Astrof\'{\i}sica de Canarias, V\'{\i}a L\'actea S/N,
           La Laguna 38200, Tenerife, Spain
       }
\date{Accepted 
      Received }

\pagerange{\pageref{firstpage}--\pageref{lastpage}}
\pubyear{2002}

\begin{document}

\maketitle

\label{firstpage}

\begin{abstract}
We present integral field spectroscopy of X2127+119, the luminous X-ray binary in the globular cluster M15, obtained with {\it INTEGRAL/WYFFOS} on the William Herschel Telescope. We find that tomograms of \heii line profiles appear to be incompatible with the previously-assumed view of X2127+119, in which the binary consists of a 1.4-\msun\space neutron star and a 0.8-\msun\space sub-giant companion near the main-sequence turn-off for M15. Our data imply a much smaller mass ratio $M_2/M_{\rm X}$ of \til0.1. We find that models of X2127+119 with black-hole compact objects give a poor fit to our data, while a neutron-star compact object is consistent with the data, implying that X2127+119's companion may have a much lower mass (\til0.1 \msun) than previously assumed. As an \til0.1-\msun\space main-sequence star would be unable to fill its Roche lobe in a binary with X2127+119's orbital period (17.1 hours), the companion is likely to be the remnant of a significantly more massive star which has had most of its envelope stripped away.
\end{abstract}


\section{Introduction}

The X-ray binary X2127+119 (AC211) in the globular cluster M15 is a very enigmatic object: despite being optically the brightest of the globular cluster XRBs, and despite almost 20 years of observations (e.g. Auri\'ere et al. 1984; Ilovaisky et al. 1987; Naylor et al. 1988; Ilovaisky 1989; Ilovaisky et al. 1993), very little is known about the system, due to the extreme difficulties associated with observing sources in the cores of globular clusters. Spectra of AC211 show blue-shifted Balmer and \hi absorption, each line often consisting of several blue-shifted absorption components at different velocities (e.g. Ilovaisky 1989), indicating a complex, structured mass outflow from the system. However, the spectra have never had the signal-to-noise, nor the orbital phase coverage, necessary to enable an understanding of the nature of AC211's mass outflows.

\begin{figure}
\begin{center}
\includegraphics[width=8.4cm]{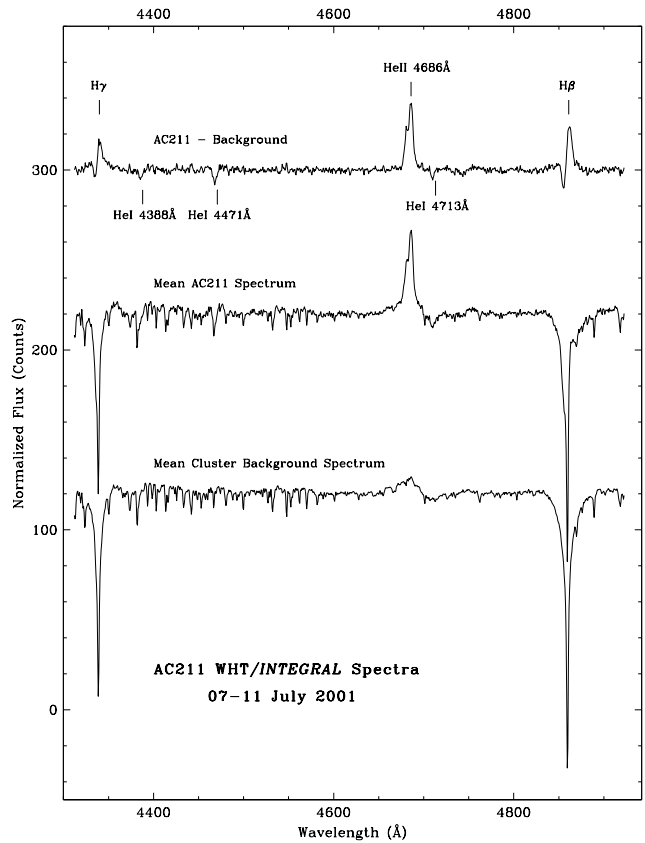}
\end{center}
\caption{Mean spectra of AC211 and the cluster background from the 2001 observing run. The top spectrum is the background-corrected mean spectrum of AC211.}\label{fig:mean}
\vspace{5mm}
\end{figure}

AC211 is an accretion disc corona (ADC) source: a high inclination system in which the compact object is permanently obscured behind the accretion disc rim, and from which we therefore detect only a small fraction of the intrinsic X-ray flux. Little is known about AC211's compact object, or about its donor star. It is assumed that the former is a neutron star rather than a black hole, on the grounds that all the other globular cluster low-mass X-ray binaries (LMXBs) contain neutron stars. Observed X-ray bursts previously assumed to have come from AC211 (e.g. Dotani et al. 1990) are now believed to have been produced by a newly-discovered second LMXB in the core of M15, \til3 arcsec from AC211 (White \& Angelini 2001), so AC211's status as a neutron-star system is no longer as certain as previously believed. From its high optical luminosity and inferred intrinsic X-ray flux, it is believed that AC211 has a high mass transfer rate (e.g. Charles, Jones \& Naylor 1986). Together with its long orbital period (17.1 hours), this has led to the assumption that the donor star must be a sub-giant which has relatively recently begun to evolve off the main sequence. The main sequence turn-off point for M15 corresponds to 0.8 \msun\space (Fahlman et al. 1985), which has therefore been assumed to be the mass of AC211's donor.

AC211 lies within an exceptionally crowded stellar field, dominated by giants significantly more luminous in the optical wave-band. As a result, ground-based spectroscopy of AC211 has been heavily contaminated by the cluster background (e.g. Naylor et al. 1988; Ilovaisky 1989). The HST/STIS is able to resolve AC211 and provide spectra free from cluster background light (Ioannou et al. 2003), but temporal coverage with the HST is poor: four to five orbits of HST spectroscopy (a typical allocation) covers only 15-20{\th}\% of AC211's 17.1-h orbital cycle.

The background-subtraction problems endemic to ground-based spectroscopy of AC211, and the impossibility of obtaining long, uninterrupted observing runs with the HST, therefore led to a desire for high-spatial-resolution two-dimensional spectroscopy of the cluster core. We believed that ground-based integral field spectroscopy would enable us to remove the cluster-core contamination from AC211's spectra, and at the same time give us much longer observing runs and greatly improved temporal coverage.

\section{Observations}

We conducted two observing runs on AC211 (3 nights in August 1999, 5 nights in July 2001) using the {\it INTEGRAL/WYFFOS} integral field spectrograph (Arribas et al. 1998; Bingham et al. 1994) on the William Herschel Telescope (WHT) on La Palma. We used a grating of 1200 g/mm and the fibre bundle SB1 which gave a 7.8 $\times$ 6.4-arcsec field of view, each fibre having a 0.45-arcsec core diameter. This set-up gave us a spectral resolution of 1.4~\AA. Table{\th}\ref{table:log} gives the log of observations. Each {\it INTEGRAL} exposure produces typically \til 200 individual spectra, which can be reconstructed to form wavelength-resolved images.

The data were reduced using tasks in the multi-fibre spectroscopy analysis package {\it noao.twodspec.apextract} in {\it iraf}, which extracts 205 spectra from each exposure and corrects them for fibre-throughput variations, removes scattered light, and performs wavelength calibration. The spectra were not flux-calibrated, but were normalized to the continuum. The resulting sets of spectra could then be used to construct surface maps of the cluster core at any wavelength in the spectral range. To remove the cluster-core contamination from AC211's spectra we used the spectra from the fibres surrounding AC211's location to interpolate and subtract the cluster-core background, essentially performing `aperture photometry' for each $\Delta \lambda$ interval of each spectrum. A discussion of background-subtraction with {\it INTEGRAL} spectra is given in Arribas, Mediavilla \& Fuensalida (1998) and references therein.

\begin{table}
\caption{AC211 {\it INTEGRAL} Log of Observations.}\label{table:log}
\begin{center}
\begin{tabular}{ccccc} \hline \hline
          &         &             &                  \\
Date & Spec. Range& Dispersion & No. & T$_{\rm exp}$/Exp. \\
        & (\AA)       & (\AA)      &  Exps & (s)     \\
        &             &      &    &         \\ \hline
        &             &      &    &         \\
19/8/99 & 4310 - 4920 & 0.65 & 21 & $\;$900 \\
20/8/99 & 4310 - 4920 & 0.65 & 27 & $\;$900 \\
21/8/99 & 4310 - 4920 & 0.65 & 27 & $\;$900 \\
        &             &           &         \\
07/7/01 & 4310 - 4920 & 0.65 & 04 &    1500 \\
08/7/01 & 4310 - 4920 & 0.65 & 07 &    1500 \\
09/7/01 & 4310 - 4920 & 0.65 & 12 &    1500 \\
10/7/01 & 4310 - 4920 & 0.65 & 07 &    1500 \\
11/7/01 & 4310 - 4920 & 0.65 & 11 &    1500 \\
          &             &      &    &                     \\ \hline \hline
\end{tabular}
\end{center}
\end{table}

\begin{table*}
\caption{Mean line velocities (with respect to M15's velocity) in 1999 and 2001.}\label{table:linevels}
\begin{center}
\begin{tabular}{cccccc} \hline \hline
% & & & & & \\
 & \hspace{11mm}  \hb \hspace{11mm}  &  \hspace{10mm} \hg \hspace{10mm}  & \hspace{5mm} \heg \hspace{5mm} & \hspace{4mm} \he \hspace{5mm} & \hspace{5mm} \heh \hspace{5mm} \\
 & km{\th}s$^{-1}$ &  km{\th}s$^{-1}$ & km{\th}s$^{-1}$ & km{\th}s$^{-1}$ & km{\th}s$^{-1}$ \\ \hline
%     & & & & & \\ \hline
%     & & & & & \\
 \hspace{2mm} 1999 \hspace{2mm}  & --306 $\pm$ 7 (a) & --317 $\pm$ 10 (a) & --161 $\pm$ 29 & --327 $\pm$ 14 & --70 $\pm$ 21 \\
     & --46 $\pm$ 15 (e) & --107 $\pm$ 17 (e) &               & --76 $\pm$ 11 & \\
%     & & & & & \\
2001 & --209 $\pm$ 9 (a) & --242 $\pm$ 13 (a) & --36 $\pm$ 42  & --108 $\pm$ 11 & --102 $\pm$ 16 \\
     & --70 $\pm$ 17 (e) & $\; $ --26 $\pm$ 17 (e) & & & \\ \hline \hline
%     & & & & & \\ \hline \hline
\end{tabular}

\vspace{6pt}
a = absorption component, e = emission component
\end{center}
%\vspace{10mm}
\end{table*}

\begin{figure*}
\hspace{1mm}\includegraphics[width=15cm]{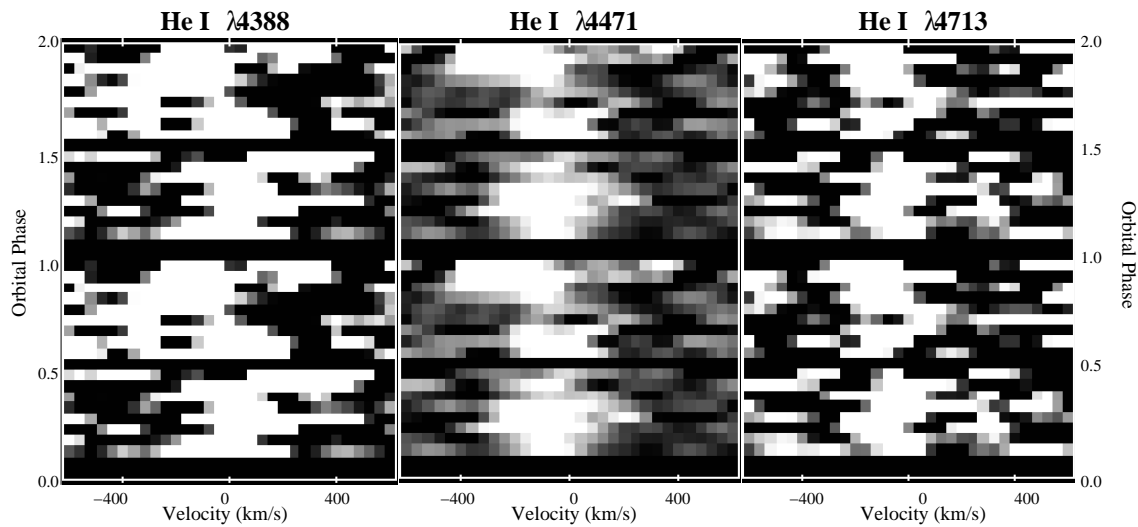}
\caption{Phase-binned, trailed spectra of AC211's \hheg, \he and \heh lines in 2001. Each spectrogram is plotted twice for clarity. The individual spectra have been normalized to the continuum. Absorption features are shown as light on a dark continuum. The flux scale is arbitrary and has been chosen to maximize the contrast between the absorption features and the continuum.}\label{fig:lines}
\vspace{3mm}
\end{figure*}

Fig.{\th}\ref{fig:mean} shows the mean AC211 and cluster background spectra from the August 2001 observing run. The top spectrum is the background-subtracted mean spectrum for AC211. The bump near 4680\thinspace\AA\space in the cluster background spectrum is an artefact of the continuum-flattening routine we used, which was chosen in such a way as to preserve any structure in the \heii wings. The same continuum-fitting parameters are used for all the background and target fibres of individual exposures, and therefore the artefact is completely removed when the background spectra are subtracted from the AC211 spectra.

The quality of the 1999 dataset is poorer compared to that of 2001. There are three reasons for this: (i) in 2001 we used longer integration times for each exposure (1500{\th}s as opposed to 900{\th}s), (ii) in 1999 the seeing was poor (\til 2 arcsec), while in 2001 it was excellent (\til 0.5-1 arcsec), and (iii) in 1999 the spectrograph's CCD detector was not properly earthed, so that an oscillating signal from the power supply was injected into each exposure as the CCD was read out, with the result that a prominent pseudo-sinusoidal pattern was incorporated into each extracted spectrum. Fortunately, the oscillation patterns in each of the 205 spectra within an individual exposure had approximately similar frequencies, amplitudes and phases, and therefore these superimposed signals could be removed reasonably successfully simply by subtracting the cluster background.

\section{Results and Discussion}

\begin{figure*}
\centering
\includegraphics[width=17cm]{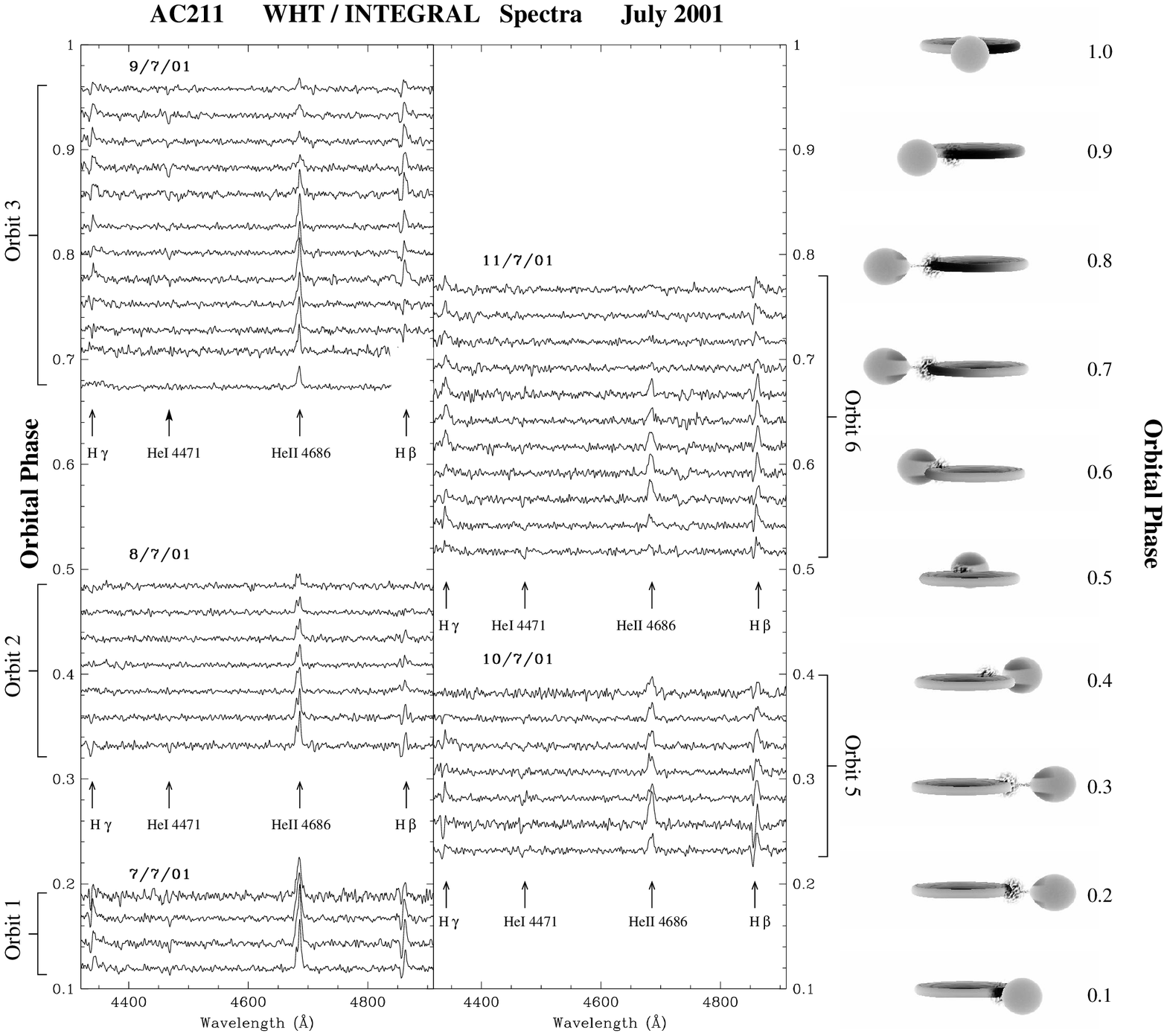}
\caption{Spectra of AC211 from the July 2001 WHT run, stacked with respect to orbital phase, and plotted alongside simple representations of AC211, in order to highlight any relationships between spectral features and components of the binary system. The observations spanned 6 binary orbits; the orbit-to-orbit variations between spectra obtained at similar orbital phases are clearly apparent.}\label{fig:phasefig}
\vspace{6mm}
\end{figure*}

Fig.{\th}\ref{fig:mean} shows that the background-subtraction technique discussed above is successful. Subtraction removes all spectral lines except those unique to AC211. The clear P Cyg profiles of AC211's \hb and \hg lines (first noted by Ilovaisky et al. 1989) are revealed, and \hheg, \he and \heh appear in absorption. Further evidence for the P Cyg profiles in the background-subtracted spectra being real rather than artefacts of the subtraction process is the clear asymmetry in the mean unsubtracted AC211 Balmer lines in Fig.{\th}\ref{fig:mean}. Because of the poorer quality of the 1999 spectra, we were unable to remove the background from the Balmer lines as successfully. However, the 1999 spectra show clear, clean \heii emission, since the background is flat in this wavelength region.

In Table \ref{table:linevels} we list the line velocities (obtained by fitting Gaussians -- double Gaussians in the case of P Cyg profiles -- to the lines) with respect to the cluster (which has a radial velocity of $-107.3 \pm 0.2$~km~s$^{-1}$; Peterson, Olszewski \& Aaronson 1986) for the 1999 and 2001 mean spectra. The short vertical bars in Fig.{\th}\ref{fig:mean} indicate the rest wavelengths (relative to the cluster) of the relevant lines; AC211's absorption lines are clearly blue-shifted with respect to the cluster velocity. These observations are consistent with results from previous observations (e.g. Naylor et al. 1988, 1989; Ilovaisky et al. 1989).

In Fig.{\th}\ref{fig:lines} we plot trailed spectrograms of the \helium lines in the 2001 dataset. In all the spectrograms and Doppler maps in this paper, the individual spectra have been normalized to the continuum and and the sense of the grayscales is such that absorption features appear light and emission features appear dark. The flux scales used are arbitrary and have been chosen to maximize the contrast between the absorption features and the continuum. In the 1999 dataset, only the \heii profiles have sufficient signal-to-noise to warrant trailing (see Fig.{\th}\ref{fig:dopp}), while in the 2001 dataset, the extreme variability of the \heii emission renders trailing meaningless. The trailed spectra in Fig.{\th}\ref{fig:lines} are phase-binned and folded on AC211's orbital period. Conclusions drawn from the spectrograms should be treated with caution: the spectra are obtained from different orbital cycles, and in Fig.{\th}\ref{fig:phasefig} we show that spectra of similar orbital phase but from different orbital cycles can be remarkably different.

In Fig.{\th}\ref{fig:phasefig} we have stacked the individual spectra from the July 2001 observations with respect to orbital phase. Alongside the spectra we have plotted simple cartoons of an X-ray binary, in order to highlight any relationships between spectral features and components of the binary system. The cartoons do not, however, show the structure of the accretion disc rim. X-ray observations (Ioannou et al. 2002) indicate that AC211's disc has a very large disc bulge which extends \til90\deg\ along the disc rim from the stream impact point along the direction of rotation. The location (but not the vertical extent) of the bulge is indicated by the darker shading on the disc edges in the cartoons. The X-ray observations also indicate that the structure of the disc bulge can vary dramatically from orbit to orbit.

\subsection{The \helium lines}

Although the \helium lines in Fig.{\th}\ref{fig:lines} are very weak, they show a clear orbital-phase dependence and a net blue-shift, which is consistent with previous observations (Naylor et al. 1988; Ilovaisky 1989). \heg appears to have a maximum blue-shift near phase 0.75; a motion normally associated with the companion star. Alternatively, \heg absorption with this motion may occur in material splashing off the stream-disc impact site (the bright spot). It should be noted, however, that the \heg line is very weak, and the apparent phase-0.75 maximum blue-shift may not be real. In fact, apart from this feature, this line appears much the same as the other \helium lines. \he and \heh have motions with maximum blue shifts occurring near phase 0.0. No obvious component within the binary's geometry moves with this motion.

In Fig.{\th}\ref{fig:phasefig} \hhe, the strongest of the \helium lines, appears strong between phases \til$0.1-0.3$ in orbital cycles 1 and 5, phases between which there is no disc-bulge obscuration and therefore where the thickness of the outer accretion disc is lowest, allowing us to see into the disc's hot inner regions. However, \he is also strong between phases \til$0.8-0.95$ in orbit 3 and between phases 0.5 and 0.6 in orbit 6, phases at which the disc bulge obscures our view of the inner disc, indicating that the disc/stream impact region may be a source of the \til 30\th000~K continuum required for \he absorption. However, it is not clear why \he should be so weak between phases \til$0.35-0.5$ in orbit 2; phases at which disc-bulge obscuration should be insignificant. From this figure it is clear that \heii can vary dramatically in strength from orbit to orbit; perhaps so too can \hhe.

\subsection{The Balmer lines}

The \hb and \hg radial velocities in Table \ref{table:linevels} seem to imply that AC211's Balmer lines are produced in an outflow more complex than a simple spherically-expanding wind: both the emission and absorption components of their P-Cyg profiles are blue shifted. This may imply that the outflow in which the Balmer lines are formed is not spherically symmetric, and that we observe more material approaching than receding. HST observations of AC211's \heu (Ioannou et al. 2003) indicate that the outflowing line-producing material exists predominantly above and below the accretion disc, with very little emitting material flowing along the orbital plane.

In Fig.{\th}\ref{fig:phasefig} the \hb and \hg P Cyg components exhibit significant changes in line depth/height and velocity. However, it is not possible to tell whether these effects are intrinsic to the line profiles, or whether they are artefacts due to over- or under-subtraction of the cluster-background Balmer profiles.

\subsection{The \heii line}

\begin{figure}
\begin{center}
\includegraphics[width=8.4cm]{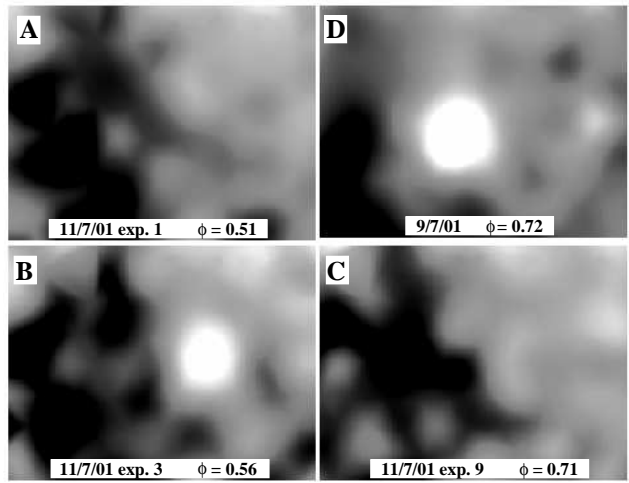}
\end{center}
\caption{A sample of \til4.5$\times$3.5-arcsec \heii images of the core of M15, demonstrating the dramatic variability of AC211's \heii flux from exposure to exposure. Panel D plots data at the same orbital phase as that in panel C, but obtained 2 days earlier, illustrating AC211's non-repeatability from orbit to orbit. (Note: observations from 9/7/01 and 11/7/01 had somewhat different pointings.)}\label{fig:hepics}
\vspace{3mm}
\end{figure}

The most prominent feature of the AC211 {\it INTEGRAL} spectra, the \heii emission line, is also the most unstable (Fig.{\th}\ref{fig:phasefig}), most notably in the 2001 data. Its flux and profile changes dramatically from orbit to orbit, and even from exposure to exposure. Fig.{\th}\ref{fig:hepics} illustrates this variability with \til4.5$\times$3.5-arcsec images of the core of M15 reconstructed from continuum-divided \heii light. In the first exposure (Fig.{\th}\ref{fig:hepics} A) AC211 has no detectable \hee flux above the continuum level; by the third (Fig.{\th}\ref{fig:hepics} B), the \hee flux is strong, but by the 9th exposure (Fig.{\th}\ref{fig:hepics} C), it has switched off again. In the final panel (Fig.{\th}\ref{fig:hepics} D), an image at the same orbital phase obtained two nights (and three orbital cycles) previously, in which the \hee flux is extremely strong and steady, is plotted for comparison.

It is interesting to note that in the exposures in which \heii is `off', the Balmer lines still show P Cyg profiles (Fig.{\th}\ref{fig:phasefig}), indicating that the Balmer- and \hhee-emitting regions do not coincide. Even more intriguingly, \he remains present in absorption, indicating that despite significant obscuration, an extremely hot (\til 30,000{\th}K) source of continuum radiation remains visible.

The orbital phases at which the \heii flux occasionally but dramatically `switches off', \til 0.5 to \til 0.7, correspond to the phases during which the region around the compact object would be obscured from our line-of-sight by the bulge in the inner part of the accretion disc created by the re-impact of an accretion stream overflowing the disc rim (see Ioannou et al. [2003] for the evidence for and a discussion of this feature in AC211's disc). It is possible that this strongly variable \heii flux is the result of `blobbiness': variations in the height or thickness of the obscuring material. Obscuration by structure in the disc seems the only plausible explanation: it may perhaps be possible to turn the \hee flux on and off by modulating the mass-transfer rate onto the compact object, but this is unlikely to produce the rapid and sharp variations in the \heii flux that we observe. In addition, the most dramatic variability occurs at phases associated either with the disc bulge resulting from the stream impact site on the disc edge or with the inner disc bulge associated with stream overflow and re-impact (however, without more extensive observations, it is not possible to know whether the variability always occurs at these phases).

\begin{figure}
\hspace{10mm}\includegraphics[width=6cm]{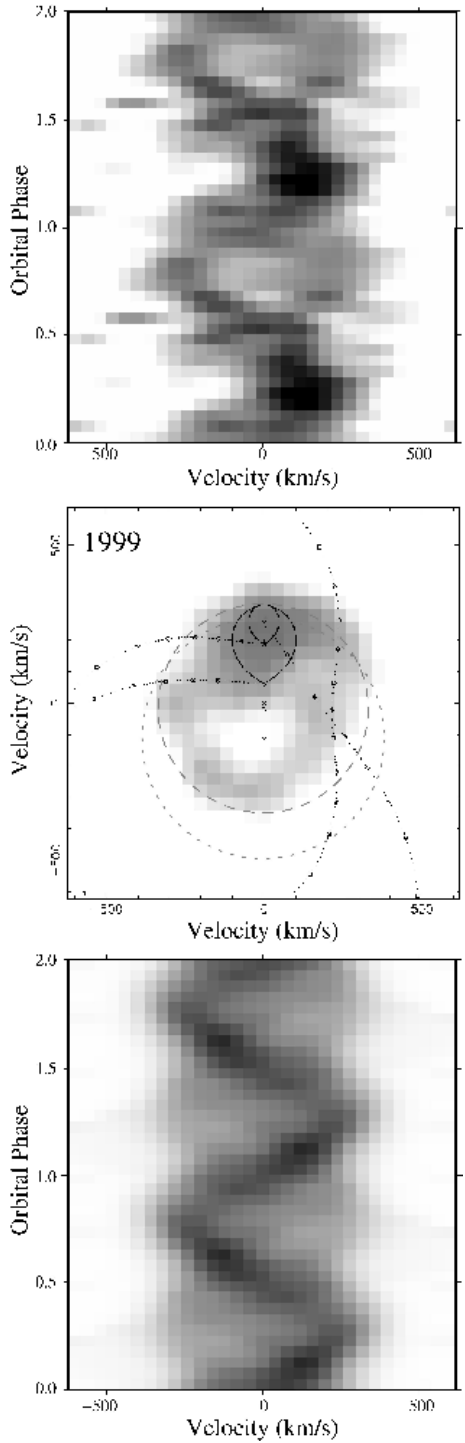}
\caption{Trailed spectra and tomogram (middle panel) of the 1999 data. Companion star Roche lobes and outer accretion disc velocities are plotted for two different mass ratios: $q = M_2/M_{\rm X} = 0.8$\msun$/1.4$\msun$ =0.57$ and $0.14$\msun$/1.4$\msun$ =0.1$. Dashed and dotted circles give the locations in velocity space of the outer disc edge ($R_{\rm disc} = 0.9 R_{\rm L_{1}}$) for $q=0.57$ (large lobe and ring) and $q=0.1$ (small lobe and ring) respectively. The location of the centre of mass of the binary system is indicated by a cross at zero velocity (upper of 3 crosses in the centre panel), and the 2 locations for the centre of mass of the compact object (for each mass ratio) are indicated by the lower two crosses. The \heii spectrogram, and the spectrogram reconstructed from the tomogram, are plotted above and below the tomogram.}\label{fig:dopp}
%\vspace{6mm}
\end{figure}

\begin{figure*}
\includegraphics[width=17cm]{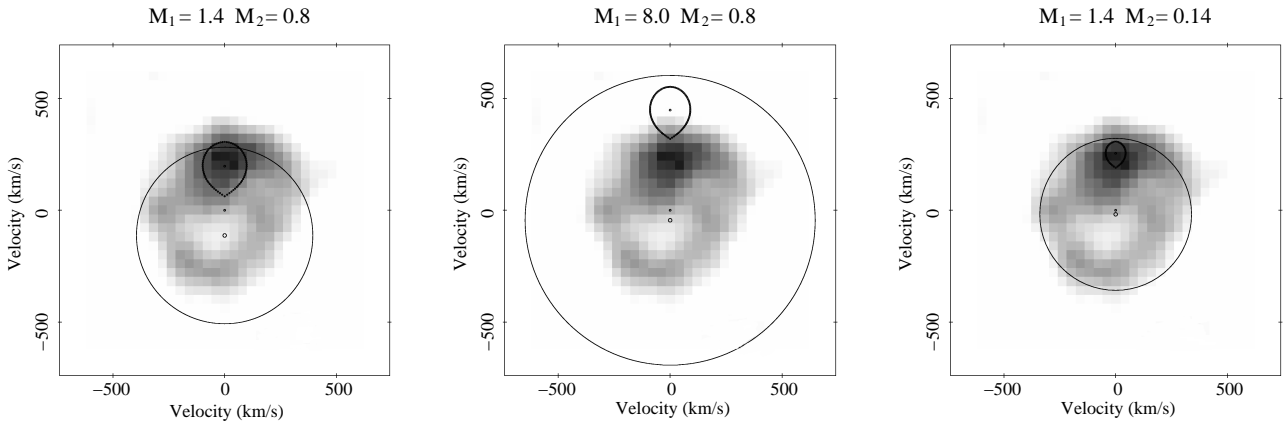}
\caption{\heii Doppler tomograms of the 1999 dataset, with companion Roche lobes and outer accretion-disc locations in velocity space overplotted for different component masses and mass ratios. In each panel, the location of the centres of mass of the binary and the compact object are indicated by a point at zero velocity and an open circle, respectively.}\label{fig:lobes}
\end{figure*}

In addition to disc-rim obscuration of the region around the compact object with respect to our line-of-sight, another possible explanation for the \heii variability may be disc-rim obscuration of the inner disc with respect to the companion star. Doppler tomograms (Figs.{\th}\ref{fig:dopp} and \ref{fig:lobes}) indicate that some of the \heii emission may be the result of X-ray reprocessing off the face of the companion star, and in \S{\th}4 we argue that there may be evidence that AC211 has an extreme mass ratio. If the latter is true, and if the height of the disc bulge varies from orbit to orbit, it is possible that the bulge may sometimes subtend a larger angle, as seen from the compact object, than the companion star, thereby shielding the latter from irradiation. This could account for the absence of some, but not all, of the \heii emission (provided that X-ray irradiation of the companion from the ADC can be ignored). 

If the variation in \heii flux is a result of disc rim or disc bulge obscuration, the rapidity of the flux variations implies that the disc rim or bulge has an unstable structure, or that the accretion stream impacting the disc bulge contains blobs of different density (not unlikely, given the evidence for `blobby' accretion stream flows in AM Her stars; e.g. Warner 1995), some causing larger increases in disc-bulge height on impact, or producing more splash, than others.

The orbit-to-orbit variation in \heii flux may also imply that the accretion disc is tilted and precesses. However, AC211's long-term X-ray light curve shows no sign of the modulations characteristic of precessing, tilted discs (Charles, Clarkson \& van Zyl 2002). In addition, we observe evidence for variable bulge obscuration in X-ray observations of AC211 (Ioannou et al. 2002), making this, rather than a tilted disc, the more likely reason for the \heii flickering.

This strong \heii variability is difficult to reconcile with what we know about AC211 from previous observations. \hee emission is understood to come from an accretion disc corona (ADC; e.g. Fabian, Guilbert \& Callanan 1987; Naylor et al. 1988), an extended region of low-density, X-ray irradiated material above and below the accretion disc. X-ray and UV observations of AC211 (Ioannou et al. 2002, 2003) show that the ADC is substantially larger than the size of the companion star; for example, in the UV the \heu line is not eclipsed (Ioannou et al. 2003).

These results indicate that the \hee emission region lies well above the accretion disc, unobscured by the companion or disc bulge throughout the orbit. The ADC is exposed to an intense X-ray flux from the compact object and inner disc; we would therefore expect \hee recombination to occur predominantly in the outer, cooler regions of the ADC, unobscured by the disc bulge, which is how the \heu emission behaves. Our \heii data indicate that either \heii recombination somehow occurs in regions of the ADC closer to the accretion disc than the \heu recombination region (assumed to be in the outer regions of the ADC), or that at the time of the 2001 {\it INTEGRAL} observations, the mass-transfer rate onto the compact object -- and therefore the accretion luminosity -- was lower. However, the XTE ASM shows no change in AC211's X-ray flux during this period.

\subsection{Trailed spectrograms, Doppler tomograms and the 
mass ratio of AC211}

In Fig.{\th}\ref{fig:dopp} we plot \heii
spectrograms, folded on orbital phase, from our 1999 dataset.  
This shows evidence for an S-wave feature
which is at greatest blue-shift near phase 0.8. 
The 2001 data shows strong variability that doesn't repeat with
the orbital cycle, and does not have complete phase coverage, 
and thus is less useful for creating a phase-resolved trail. 

Doppler tomography can be a helpful tool in interpreting trailed
spectra, thus Fig.{\th}\ref{fig:dopp} also contains a Doppler tomogram
obtained using the maximum-entropy method (Marsh \& Horne 1988).
Again, we only show this for the 1999 data, where the trail shows
similar flux levels over the orbital cycle.  The strong variability
of the 2001 data violates an assumption of Doppler tomography, that
there are no changes in visibility over the cycle. 

The S-wave feature maps to a bright region in the tomogram near
velocity coordinates (0, 200) km s$^{-1}$. Underlying this is
a diffuse ring of emission with a radius of $\approx$\, 300 km $^{-1}$.

The reconstructed trail (Fig.{\th}\ref{fig:dopp}, bottom panel) compares well with the observed trail, suggesting the Doppler map is a reasonable representation of the orbital average \heii emission distribution.

The diffuse ring in the Doppler map can be associated with accretion disc emission. 
Another possibility would be emission from a wind, which, owing to
its out-of-plane velocity, would appear `defocussed' in the tomogram.
However, if this were the case, the trail spectrogram would show
a strong asymmetry, equal to the velocity of the ring, which is not seen.  

If the ring feature results from a disc, the ring radius corresponds
to the velocity of the outer disc edge, and the ring centroid should
be located at the orbital velocity of the compact object.  Also,
an S-wave feature normally results from emission either from the
irradiated secondary star, or from the disc edge where it is struck
by the accretion stream. 

We have overplotted on the tomogram the Keplerian disc velocities and
secondary Roche lobes for the two mass ratios $q = 0.57$ and $q = 0.1$,
and assuming that the compact-star is a neutron star with a 
canonical mass of 1.4\,\msun.  The value $q = 0.57$ then corresponds
to a companion mass of 0.8 \msun, which is M15's main sequence
turn-off mass. The second value, $q = 0.1$ is then chosen as
a much-smaller value, appropriate to a \til~0.1~\msun\ companion.

The tomogram argues against the larger mass ratio of $q = 0.57$ since
the observed ring feature is clearly offset from the expected
location, and is not concentric with the Roche lobe of the primary
(see Fig.~6, left panel).

On the other hand, the observed feature is concentric with the
primary's Roche lobe for a significantly smaller mass ratio near $q
\sim 0.1$ (see Fig.~6, middle panel and right panel, which have
$q = 0.1$ but different primary masses).

With the small mass ratio but high primary mass ($M_{1}$ = 8\,\msun;
Fig.~6, middle panel) the ring velocities are much lower than
would be expected for disc emission. In particular, the S-wave feature
is at too low a velocity to be associated with the secondary.

However, with the small mass ratio and a smaller primary mass of
$M_{1} = 1.4$\,\msun\ (Fig.~6, right panel), the S-wave is then
coincident with the secondary star, and so is compatible with an
origin on the irradiated secondary.  Further, the velocities of the
ring are closer to the expected Keplerian velocities of a disc.  In
fact, the ring velocity is somewhat lower than the expected Keplerian
velocity. However, sub-Keplerian outer disc velocities (whether real
or a consequence of an incomplete understanding of how lines form in
discs) are not uncommon: the phenomenon has been observed in several
systems, for example Z Cha, whose outer disc velocity is sub-Keplerian
by \til 30\% (Wade \& Horne 1988), and X1822--371 (Harlaftis, Charles
\& Horne 1997).

Thus, despite the limitations of Doppler tomography in data of limited
quality, the tomograms suggest that the mass ratio of AC211 is low ($q
\approx 0.1$) and that the compact-star mass is nearer that of a
neutron star (1.4\,\msun) than a typical black-hole mass (8\,\msun).

\section{The evolutionary status of AC211}

The tomography presented in Fig.{\th}\ref{fig:lobes} argues against the companion mass
being near the 0.8 \msun\ turn-off mass of M15, instead suggesting
that the compact object is a neutron star, and the companion mass
as low as $\approx$\,0.1 \msun. 

It has in the past been assumed (e.g.\ Naylor et al. 1988; Bailyn, Garcia
\& Grindlay 1989) that AC211's high luminosity, and by implication
high mass transfer rate, has been fuelled by the expanding envelope of
a companion evolving off the main sequence. A companion with a mass
much lower than the main-sequence turn-off mass therefore leads to two
difficulties: (i) a lower-mass star would not yet have started
evolving off the main sequence and would therefore be unable to
fill its Roche-lobe in a binary with a 17.1-h orbital period, and (ii) a low-mass companion would give a mass-transfer rate too low 
(e.g.\ Pringle 1985; Edwards \& Pringle 1987; Podsiadlowski, Rappaport
\& Pfahl 2002) to account for AC211's luminosity.

Both these difficulties could be solved if the companion was once a much
more massive star which has had most of its envelope stripped
off. Evolved companion models, in which nuclear evolution drives the
mass transfer, have been developed by Taam (1983) and Webbink,
Rappaport \& Savonije (1983). All the properties of the companion star
depend on the mass $M_{\rm c}$ of its helium core, and not on its
total mass $M_2$. Hydrogen shell burning increases $M_{\rm c}$, which
in turn increases the radius of the companion's envelope, which drives
the mass transfer. $M_{\rm c}$ will increase, and the companion's
envelope will be stripped away through Roche lobe overflow, until
$M_{\rm c}=M_2$, at which point the companion loses its envelope and
detaches from its Roche-lobe, becoming a low mass helium white dwarf.

King (1993) has shown that the companion star in the black hole candidate system V404 Cyg conforms to a stripped-giant model. By inserting AC211's orbital period into King's equations, we obtain minimum-mass and maximum-mass solutions of 0.15 and 0.98~\msun\ respectively for $M_2$, and 0.15 and 0.17~\msun\ respectively for $M_{\rm c}$. (The minimum-mass solution is found by assuming that $M_{\rm c}=M_2$, i.e. that the companion has had its envelope removed, while the maximum-mass solution is found by assuming the Sch\"onberg-Chandrasekhar limiting mass for $M_{\rm c}$ of 0.17~$M_2$, below which the companion would not have left the main sequence.) The results from our Doppler maps appear to be consistent with the lower end of this range, where $M_2$ is close to $M_{\rm c}$, which would imply that the companion has had most of its envelope stripped away.

There is no reason why a very low mass, stripped-giant companion could
not provide an accretion rate sufficient to power AC211's
luminosity. The luminosity intercepted by the companion star may be
several orders of magnitude greater than its internal luminosity
(Phillips \& Podsiadlowski 2002), and the change in the entropy in the
layer where the irradiation flux is thermalized may drastically change
the equilibrium structure of the star, which may lead to significant
expansion (Podsiadlowski 1991). A mass ratio of \til\,0.1 would mean
that the companion is shielded from the compact object by the disc
rim; however, it would still be exposed to X-ray irradiation from the
ADC. If the ADC scatters a fraction of only $\sim 10^{-4} - 
10^{-3}$ of the system's X-ray flux onto the companion, that would be
enough to completely change its structure and drive up the mass
transfer rate (Podsiadlowski 2001, private
communication). Irradiation-enhanced mass transfer is a significant
possibility in AC211: our \heii Doppler maps indicate that the
companion is indeed being irradiated.

Although the results of our tomograms need to be treated with caution considering the quality of our data and our limited spectral and temporal resolution, we nevertheless believe that these results point to a significantly different mass-ratio for the system, and therefore provide us with a valuable insight into the evolutionary state of AC211.

% However, if AC211's high mass transfer rate is a consequence of irradiation of the companion by the ADC, we would need to find an explanation for why this does {\it not} happen in other globular cluster LMXBs, which all appear to have lower mass transfer rates.

%\section{Conclusions}

%This paper presents WHT/{\it INTEGRAL} spectroscopy of AC211 from two observing runs in 1999 and 2001. The most significant result from this work is the possibility that we have constrained AC211's component masses and mass ratio to 1.4 \msun, \til 0.1 \msun{\space}and \til 0.1, respectively.

%We present \heii Doppler maps which, unusually, appear to exhibit emission features associated with the accretion disc. The Balmer lines show strong P Cyg profiles with emission components bluer than those expected from spherically symmetrical outflows, perhaps indicating an origin in a bipolar accretion disc wind.

%The \heii flux shows strong flickering between phases 0.5 and 0.7 (but only during some orbits), indicating variable obscuration of the \hee line-forming region by the disc bulge. This appears to contradict results obtained from HST UV spectra, which indicate that the \hee line-forming region is located high above the disc plane, and should therefore remain unobscured.

\section*{Acknowledgments}

This paper is based on observations made with the William Herschel Telescope operated on the island of La Palma by the Isaac Newton Group in the Spanish Observatorio del Roque de los Muchachos of the Instituto de Astrofisica de Canarias.

LvZ would like to thank Rob Hynes for the use of his {\sc binsim} binary visualization code, Philipp Podsiadlowski for useful discussions, and Tom Marsh for the use of his {\sc molly, doppler} and {\sc trailer} spectral analysis software. LvZ acknowledges the support of scholarships from the Vatican Observatory, the National Research Foundation (South Africa), the University of Cape Town, and the Overseas Research Studentship scheme (UK).

TN acknowledges the support of a PPARC Advanced Fellowship during part of this work.

We would like to thank the anonymous referee for helpful comments.

\label{lastpage}

\end{document}